\documentclass[conference]{IEEEtran}
\IEEEoverridecommandlockouts
\usepackage{preamble}
\usepackage{balance}
\IEEEoverridecommandlockouts
\IEEEpubid{\makebox[\columnwidth]{Accepted at COMSNETS 2026 Graduate Forum (Best Paper Runner-Up) \hfill} \hspace{\columnsep}\makebox[\columnwidth]{ }}

\begin{document}

\title{MCP-Diag: A Deterministic, Protocol-Driven Architecture for AI-Native Network Diagnostics}

\author{\IEEEauthorblockN{Devansh Lodha, Mohit Panchal and Sameer G. Kulkarni}
\IEEEauthorblockA{\textit{Indian Institute of Technology Gandhinagar}}}
\maketitle

\begin{abstract}
The integration of Large Language Models (LLMs) into network operations (AIOps) is hindered by two fundamental challenges: the \textit{stochastic grounding problem}, where LLMs struggle to reliably parse unstructured, vendor-specific CLI output, and the \textit{security gap} of granting autonomous agents shell access. This paper introduces MCP-Diag, a hybrid neuro-symbolic architecture built upon the Model Context Protocol (MCP). We propose a deterministic translation layer that converts raw \texttt{stdout} from canonical utilities (\texttt{dig}, \texttt{ping}, \texttt{traceroute}) into rigorous JSON schemas before AI ingestion. We further introduce a mandatory ``Elicitation Loop" that enforces Human-in-the-Loop (HITL) authorization at the protocol level. 
Our preliminary evaluation demonstrate that MCP-Diag achieving $100\%$ entity extraction accuracy with less than $0.9\%$ execution latency overhead and $3.7\times$ increase in context token usage.

%Quantitative evaluation against the top 500 global domains demonstrates the architectural trade-offs: while the protocol introduces a $3.7\times$ increase in context token usage due to schema enforcement, it incurs a negligible execution latency overhead of $<0.9\%$ and achieves 100\% entity extraction accuracy, effectively eliminating the stochastic parsing failures observed in baseline agents. We demonstrate the system's capabilities through a series of use cases, showcasing a secure, production-ready framework for advanced network management.

\end{abstract}

\section{Introduction}
\label{sec:intro}
The escalating complexity of modern network infrastructure, characterized by distributed microservices and dynamic, software-defined topologies, demands a paradigm shift in network management \cite{sdn_challenges_2024}. Traditional manual approaches where operators type commands into SSH terminals are no longer scalable and are prone to human error \cite{edelman2018network, rfc3535}. This has given rise to AIOps, utilizing Large Language Models (LLMs) to automate diagnostics and operational workflows \cite{zhang2025survey, liu2024large}. However, integrating LLMs with essential command-line utilities is impeded by two fundamental architectural barriers: the ``Translation Gap" and the ``Governance Gap."

The `Translation Gap' arises because utilities such as \texttt{ping}, \texttt{traceroute}, and \texttt{dig} were designed for human interpretation, not machine consumption. Their output is unstructured and highly variable; for instance, the textual output of \texttt{ping} differs subtly between \texttt{iputils} (Linux) and BSD-derived systems (macOS), and handles timeout representations differently \cite{jc_library}. Relying on an LLM to ``screen-scrape" this data is inherently probabilistic \cite{hong2025comprehensive}. An agent might correctly identify latency in one session but hallucinate an IP address in the next.

The `Governance Gap' is even more critical. Granting an LLM direct shell access via standard \texttt{subprocess} libraries poses an immense security risk. Malicious actors could exploit prompt injection attacks to trick the AI into executing arbitrary commands \cite{owasp_llm_top10_2025, cve_2025_1753}. Even without malicious intent, a stochastic agent stuck in a reasoning loop could ``hallucinate" a destructive command (e.g., \texttt{rm -rf /}) or accidentally trigger a Denial of Service (DoS) by spawning thousands of concurrent probes. A production-grade framework must enforce the principle of least privilege at the protocol level, not just via ``system prompts" that can be bypassed.
We propose a solution built upon the Model Context Protocol (MCP)~\cite{mcp_spec}. We introduce \textbf{MCP-Diag}, a functional implementation and architectural validation of MCP for network diagnostics. Our system bridges these gaps via three key mechanisms:
\begin{itemize}
    \item \textbf{Deterministic Grounding:} We inject a middleware layer (powered by the \texttt{jc} library \cite{jc_library}) that intercepts raw \texttt{stdout} and serializes it into rigorous, type-safe JSON schemas. This ensures the LLM reasons over structured data objects, eliminating parsing hallucinations.
    \item \textbf{Protocol-Enforced Governance:} We utilize MCP's \texttt{elicitation} primitive to make tool execution impossible without an explicit, cryptographic-like handshake from the human operator.
    \item \textbf{Hybrid Transport:} We introduce a dual-channel communication model, using standard MCP transports for control logic and Server-Sent Events (SSE) for asynchronous data streaming, preventing agent timeouts during long-running convergence tests.
\end{itemize}

%\balance
% Architecture Diagram
% --- Architecture Diagram (Spanning Both Columns) ---
\begin{figure}[!t]
    \centering
    \includegraphics[width=\columnwidth]{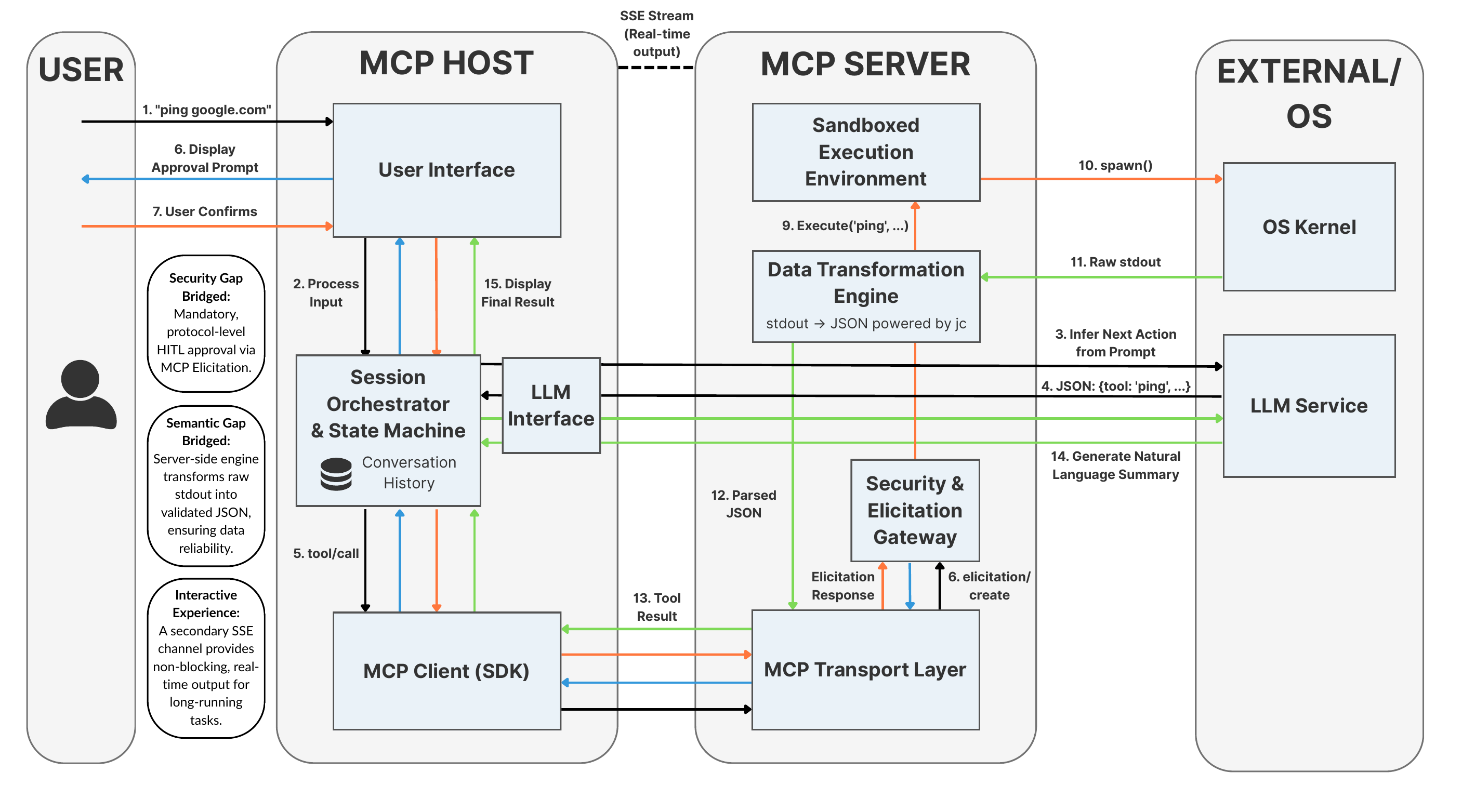}
    %\vspace{-4mm}
    \caption{MCP-Diag Architecture: The MCP Host (left) orchestrates the flow between the user, the LLM, and the MCP Server (right). Critically, all tool execution is centralized on the Server, which enforces governance via a mandatory MCP Elicitation loop (blue arrows) back to the Host, and guarantees reliable data by transforming raw stdout into validated JSON.}
    \label{fig:architecture}
    %\vspace{-6mm}
\end{figure}

\section{Related Work}
\label{sec:related_work}
The landscape of LLM-integrated operations tools (summarized in Table \ref{tab:comparison}) is characterized by a trade-off between generality and reliability. General agentic frameworks such as LangChain \cite{langchain} and Open Interpreter \cite{openinterpreter} offer high flexibility but suffer from high hallucination risks due to probabilistic parsing of raw text. AI Terminals like Shell-GPT \cite{shellgpt} and Gorilla CLI \cite{gorilla_cli} introduce ``User-Gated" safety, where commands require manual execution, yet they lack the stateful session orchestration necessary for autonomous remedial workflows (e.g., ``Trace the path only if latency exceeds 50ms"). 

Early implementations in the Model Context Protocol ecosystem, such as \texttt{mcp-nettools} \cite{mcp_nettools_github}, act as simple wrappers that pass raw binary output directly to the LLM. MCP-Diag is the first architecture to implement a \textit{Secure Translation Layer}. Unlike existing solutions, we utilize \textit{Deterministic Output Parsing} (via \texttt{jc} schemas) to ensure the LLM reasons over verified data objects. Furthermore, our custom client extends the protocol to solve the ``Blocking Wait" problem inherent in standard MCP tool calls by utilizing a side-channel SSE stream for real-time diagnostic feedback.

\begin{table*}[t]
\centering
\caption{Architectural Comparison of LLM-Driven Operational Tools}
\label{tab:comparison}
\resizebox{\textwidth}{!}{%
\begin{tabular}{lccccccc}
\toprule
\textbf{System} & \textbf{Category} & \textbf{Output Parsing} & \textbf{Hallucination Risk} & \textbf{Execution Safety} & \textbf{Human-in-the-Loop} & \textbf{State Awareness} & \textbf{Streaming} \\ 
\midrule
\textbf{mcp-nettools} \cite{mcp_nettools_github} & MCP Wrapper & \begin{tabular}[c]{@{}c@{}}Probabilistic\\ (Raw Text)\end{tabular} & High & \begin{tabular}[c]{@{}c@{}}Host-Dependent \end{tabular} & \begin{tabular}[c]{@{}c@{}}None\\ (Pass-through)\end{tabular} & \begin{tabular}[c]{@{}c@{}}Stateless\\ (One-off calls)\end{tabular} & \begin{tabular}[c]{@{}c@{}}Blocking\\ (Wait for Exit)\end{tabular} \\ 
\cmidrule{1-8}
\textbf{LangChain} \cite{langchain} & Toolkit & \begin{tabular}[c]{@{}c@{}}Probabilistic\\ (Regex/Raw)\end{tabular} & High & \begin{tabular}[c]{@{}c@{}}Client-Side\\ (Optional ``y/n" Prompt)\end{tabular} & \begin{tabular}[c]{@{}c@{}}None\\ (Pass-through)\end{tabular} & \begin{tabular}[c]{@{}c@{}}Window-Based\\ (Token limit)\end{tabular} & \begin{tabular}[c]{@{}c@{}}Blocking\\ (Wait for Exit)\end{tabular} \\ 
\cmidrule{1-8}
\textbf{Open Interpreter} \cite{openinterpreter} & General Agent & \begin{tabular}[c]{@{}c@{}}Probabilistic\\ (Raw Text)\end{tabular} & High & \begin{tabular}[c]{@{}c@{}}Isolation Only\\ (Net. Exposed)\end{tabular} & \begin{tabular}[c]{@{}c@{}}Chat-Based\\ (Can Bypass)\end{tabular} & \begin{tabular}[c]{@{}c@{}}Stateful Loop\\ (Preserves CWD)\end{tabular} & \begin{tabular}[c]{@{}c@{}}Buffered\\ (Piped Output)\end{tabular} \\ 
\cmidrule{1-8}
\textbf{Shell-GPT} \cite{shellgpt} & CLI Assistant & \begin{tabular}[c]{@{}c@{}}Probabilistic\\ (Raw Text)\end{tabular} & \begin{tabular}[c]{@{}c@{}}Medium\\ (Suggestion)\end{tabular} & \begin{tabular}[c]{@{}c@{}}User-Gated\\ (Flag Prompt)\end{tabular} & \begin{tabular}[c]{@{}c@{}}Enforced\\ (User Execs)\end{tabular} & \begin{tabular}[c]{@{}c@{}}Session-Based\\ (Cache File)\end{tabular} & \begin{tabular}[c]{@{}c@{}}Real-Time\end{tabular} \\ 
\cmidrule{1-8}
\textbf{Warp AI} \cite{warp_ai} & AI Terminal & \begin{tabular}[c]{@{}c@{}}Probabilistic\\ (Raw Text)\end{tabular} & \begin{tabular}[c]{@{}c@{}}Medium\\ (Suggestion)\end{tabular} & \begin{tabular}[c]{@{}c@{}}User-Gated\\ (CMD+Enter)\end{tabular} & \begin{tabular}[c]{@{}c@{}}Enforced\\ (User Execs)\end{tabular} & \begin{tabular}[c]{@{}c@{}}Terminal Context\\ (Shell History)\end{tabular} & \begin{tabular}[c]{@{}c@{}}Real-Time\\ (Native UI)\end{tabular} \\ 
\cmidrule{1-8}
\textbf{Gorilla CLI} \cite{gorilla_cli} & Cmd Generator & \begin{tabular}[c]{@{}c@{}}Probabilistic\\ (Raw Text)\end{tabular} & \begin{tabular}[c]{@{}c@{}}Low\\ (Finetuned Model)\end{tabular} & \begin{tabular}[c]{@{}c@{}}User-Gated\\ (User Execs)\end{tabular} & \begin{tabular}[c]{@{}c@{}}Enforced\\ (User Execs)\end{tabular} & \begin{tabular}[c]{@{}c@{}}Stateless\\ (One-off Cmds)\end{tabular} & \begin{tabular}[c]{@{}c@{}}Real-Time \end{tabular} \\ 
\cmidrule{1-8}
\textbf{Anthropic Comp. Use} \cite{anthropic_computer_use} & UI Agent & \begin{tabular}[c]{@{}c@{}}Probabilistic\\ (Visual OCR)\end{tabular} & \begin{tabular}[c]{@{}c@{}}High\\ (Visual Errors)\end{tabular} & \begin{tabular}[c]{@{}c@{}}Isolation Only\\ (Net. Exposed)\end{tabular} & \begin{tabular}[c]{@{}c@{}}High Friction / None\\ (Observation)\end{tabular} & \begin{tabular}[c]{@{}c@{}}Stateful\\ (Visual History)\end{tabular} & \begin{tabular}[c]{@{}c@{}}Real-Time\\ (Visual Proc.)\end{tabular} \\ 
\midrule
\textbf{MCP-Diag (This Work)} & \textbf{MCP Tool} & \begin{tabular}[c]{@{}c@{}}\textbf{Deterministic}\\ (\texttt{jc} Schema)\end{tabular} & \begin{tabular}[c]{@{}c@{}}\textbf{Very Low}\\ (JSON, Validated)\end{tabular} & \begin{tabular}[c]{@{}c@{}}\textbf{Protocol-Level}\\ (Mandatory Elicit.)\end{tabular} & \begin{tabular}[c]{@{}c@{}}\textbf{Enforced}\\ (Blocking State)\end{tabular} & \begin{tabular}[c]{@{}c@{}}\textbf{Stateful Session}\\ (Orchestrated)\end{tabular} & \begin{tabular}[c]{@{}c@{}}\textbf{Real-Time}\\ (Hybrid SSE)\end{tabular} \\ 
\bottomrule
\end{tabular}%
}
\end{table*}

\section{System Architecture}
\label{sec:architecture}
The MCP-Diag system uses decoupled client-server model and adheres to the open Model Context Protocol standard. The design follows the ``Sidecar Pattern," where the diagnostic logic is isolated from the inference engine. The architecture comprises two primary components: the MCP Server, which acts as a deterministic execution gateway, and the MCP Host, which functions as the stateful session orchestrator.

\subsection{Design and Execution Model}
The control flow, as visualized in \cref{fig:architecture}, proceeds through a ``U-shaped" lifecycle is described below.
\newline\noindent\textbf{1. AI-Driven Plan Generation (Steps 1-5):} The lifecycle begins when the user provides a natural language query (Step 1). The Session Orchestrator appends this to the history (Step 2) and queries the LLM (Step 3). The LLM returns a structured tool call plan (Step 4), which the Host sends to the Server as a formal \texttt{tool/call} request (Step 5).
\newline\noindent
\textbf{2. Secure, Server-Side Execution (Steps 6-12):} %The MCP Server does not immediately execute the request. Instead, 
The Security Gateway of MCP Server initiates a mandatory Elicitation loop (Steps 6-8). On receiving user confirmation, it dispatches the command (Steps 9-10). 
The resulting raw \texttt{stdout} is piped to the Data Transformation Engine (Step 11), which uses the \texttt{jc} utility to parse it into a structured JSON object (Step 12).
\newline\noindent
\textbf{3. Synthesis and Response (Steps 13-15):} The Server returns the clean JSON to the Host (Step 13). The LLM synthesizes this structured data into a human-readable summary (Step 14), completing the loop (Step 15).

\subsection{Secure Data Transformation Pipeline}
To resolve the ``Translation Gap," the Server implements a strict input-output sanitization pipeline. Unlike traditional agent frameworks that expose a raw pseudo-terminal (PTY) to the LLM, MCP-Diag enforces a rigid data contract.
The execution flow is formalized in \cref{alg:secure_execution}.

\begin{algorithm}[t]
\footnotesize
\caption{Subprocess IPC and Schema Enforcement}
\label{alg:secure_execution}
\begin{algorithmic}[1]
\Require ToolRequest $R = \langle \text{Cmd } C, \text{Args } A \rangle$
\Ensure Validated JSON Object $J \in \text{Schema}(C)$
\Procedure{HandleToolCall}{$R$}
    \State $\mathcal{M} \leftarrow$ Payload(``Authorize execution of: '' $+ C$)
    \State $\tau_{\text{auth}} \leftarrow$ ELICIT\_HANDSHAKE($\mathcal{M}$) \Comment{Protocol-level HITL}
    \If{$\tau_{\text{auth}}.\text{action} \neq$ ACCEPT} 
        \State \Return \texttt{Err}(AuthDenied) 
    \EndIf
    
    \State $P_{\text{diag}} \leftarrow$ Spawn($C, A$) \Comment{Process 1: Diagnostic Tool}
    \State $P_{\text{parse}} \leftarrow$ Spawn(\texttt{``jc''}, [\texttt{``--''}$+C$]) \Comment{Process 2: Serializer}
    
    \State $\Phi(P_{\text{diag}}.\text{stdout} \to P_{\text{parse}}.\text{stdin})$ \Comment{Atomic Stdio Pipe}
    \State $\Lambda(P_{\text{diag}}.\text{stdout} \to \text{SSE\_Buffer})$ \Comment{Async Client Feed}
    
    \State Wait($P_{\text{diag}}, P_{\text{parse}}$)
    
    \State $\epsilon \leftarrow$ ExitCode($P_{\text{diag}}$)
    \If{$\epsilon \in \text{SuccessSet}(C)$} \Comment{Tool-specific exit validation}
        \State $J \leftarrow$ Deserialize($P_{\text{parse}}.\text{stdout}$)
        \State \Return $J$
    \Else 
        \State \Return \texttt{Err}(\text{RuntimeError}, $P_{\text{diag}}.\text{stderr}$) 
    \EndIf
\EndProcedure
\end{algorithmic}
\end{algorithm} 

When a tool call is received, the system spawns two coupled processes: the diagnostic tool (e.g., \texttt{ping}) and a serialization engine (the \texttt{jc} library). The standard output (\texttt{stdout}) of the diagnostic tool is piped directly to the standard input (\texttt{stdin}) of the serializer. 
This ensures that the LLM never interacts with unstructured text; it only receives verified, schema-compliant JSON objects.

\subsection{Protocol-Enforced Governance}
To address the ``Security Gap," MCP-Diag rejects the industry-standard approach of client-side confirmation dialogs. Instead, we implement governance at the protocol level. The Server's state machine defaults to a \texttt{BLOCK} state for all sensitive network operations. Upon receiving a request, the Server suspends the execution thread and emits an \texttt{elicitation} message. The operation remains atomically locked until a valid signed confirmation token is received from the Host.

\textit{Security Inheritance:} It is crucial to note that by architecting strictly atop the Model Context Protocol, MCP-Diag inherits the security posture of the standard itself. 
\hide{As the MCP specification evolves to address vulnerabilities—such as the recent hardening of capability negotiation and transport-level encryption—our diagnostic layer automatically absorbs these improvements without requiring application-level refactoring.
}
This shared security responsibility model significantly reduces the attack surface compared to bespoke agent frameworks.

\subsection{Hybrid Transport Layer}
Network diagnostics introduce a unique concurrency challenge: tools like \texttt{ping} are long-running processes that can exceed standard RPC timeouts. We resolve this by implementing a Hybrid Transport Architecture:
\begin{itemize}
    \item \textbf{Control Plane (Synchronous):} All atomic operations—handshakes, capability negotiation, and elicitation requests—occur over the standard MCP transport (Stdio/HTTP).
    \item \textbf{Data Plane (Asynchronous):} For streaming tools, the Server opens a secondary side channel using Server-Sent Events (SSE). The Host subscribes to this stream, allowing the Server to push real-time \texttt{stdout} frames immediately as they occur.
\end{itemize}

\section{System Validation}
\label{sec:validation}
\begin{figure}[htbp]
    \centering
    \includegraphics[width=\columnwidth]{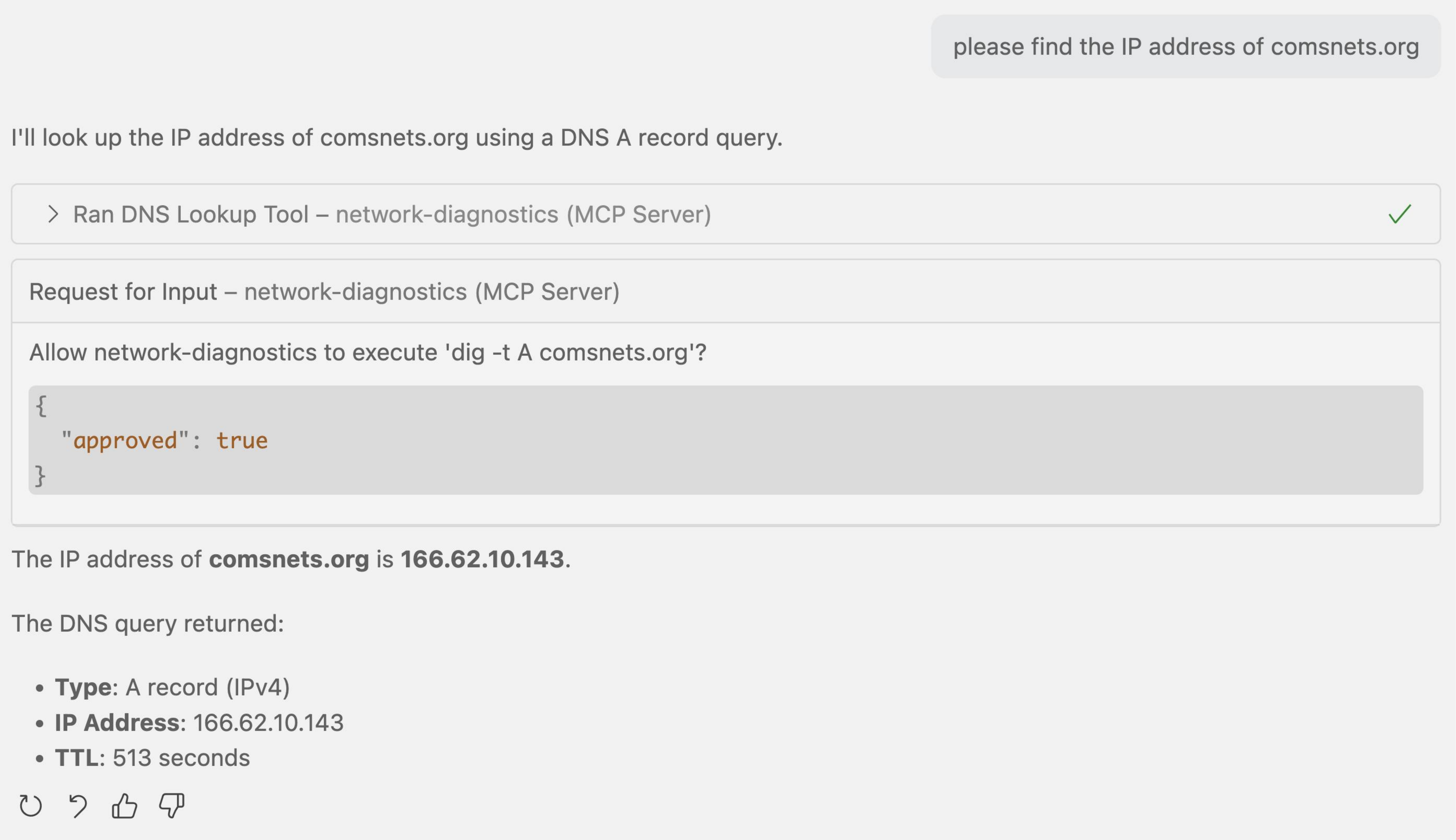}
    %\vspace{-3mm}
    \caption{%\textbf{System Validation:} 
    MCP-Diag (server) running in VS Code (Host). %, proving protocol compliance and elicitation support.
    }
    %\vspace{-4mm}
    \label{fig:use_case_security}
\end{figure}

We validated the protocol adherence of the MCP-Diag server using a standard, third-party MCP host (VS Code MCP Host). As illustrated in Fig. \ref{fig:use_case_security}, the server strictly enforces the mandatory elicitation handshake for sensitive operations, ensuring compliance with the JSON-RPC 2.0 based MCP specification.

This functional test confirms that the architecture is compatible with any compliant MCP host. While the core MCP transport logic is strictly maintained, our custom client implements an optional SSE side-channel extension for real-time, line-by-line streaming. This extension enables low-latency feedback for long-running diagnostic tasks without violating the request-response cycle of the primary MCP transport.

\section{Performance Evaluation}
\label{sec:evaluation}

The complete source code for the MCP-Diag server implementation, the custom client, and the automated benchmark suite has been made publicly available \cite{mcp_diag_repo}.

To quantify the architectural trade-offs of deterministic grounding, we conducted a large-scale benchmark ($N=500$ trials) targeting the global Top 500 domains \cite{moz_top500}. The benchmark task required the agent to execute a \texttt{traceroute} to a target domain and extract the third-hop IP address.

\subsection{Benchmark Configuration and Methodology}
We compared two distinct architectural patterns to isolate the impact of protocol-driven diagnostics:

\begin{itemize}
    \item \textbf{Baseline:} A standard agentic workflow where the prompt is sent to the LLM, which generates a raw CLI command string. The system parses this string using regex, executes it via a direct \texttt{subprocess} call, and returns the unstructured \texttt{stdout} to the LLM for probabilistic entity extraction.
    \item \textbf{MCP-Diag (Benchmark Mode):} The system executes tools via the MCP tool-call primitive. For this study, the server was configured in a ``Headless'' mode with the SSE side-channel and the mandatory Elicitation loop disabled. This configuration isolates the computational latency of serialization, transport, and schema validation from human reaction time.
\end{itemize}

% \subsubsection{Inference and Execution Environment}
All trials utilized the Gemma-3-27b-it model accessed via the Google AI API. The client-side orchestration and MCP server were executed within a local Bun/V8 JavaScript runtime. To ensure statistical validity, we enforced a JIT Warmup (preliminary execution to stabilize the runtime) and Path Priming (unmeasured traceroute before every trial to populate intermediate hop ARP tables and routing caches).

\begin{figure*}[t]
    \centering
    \includegraphics[width=\textwidth]{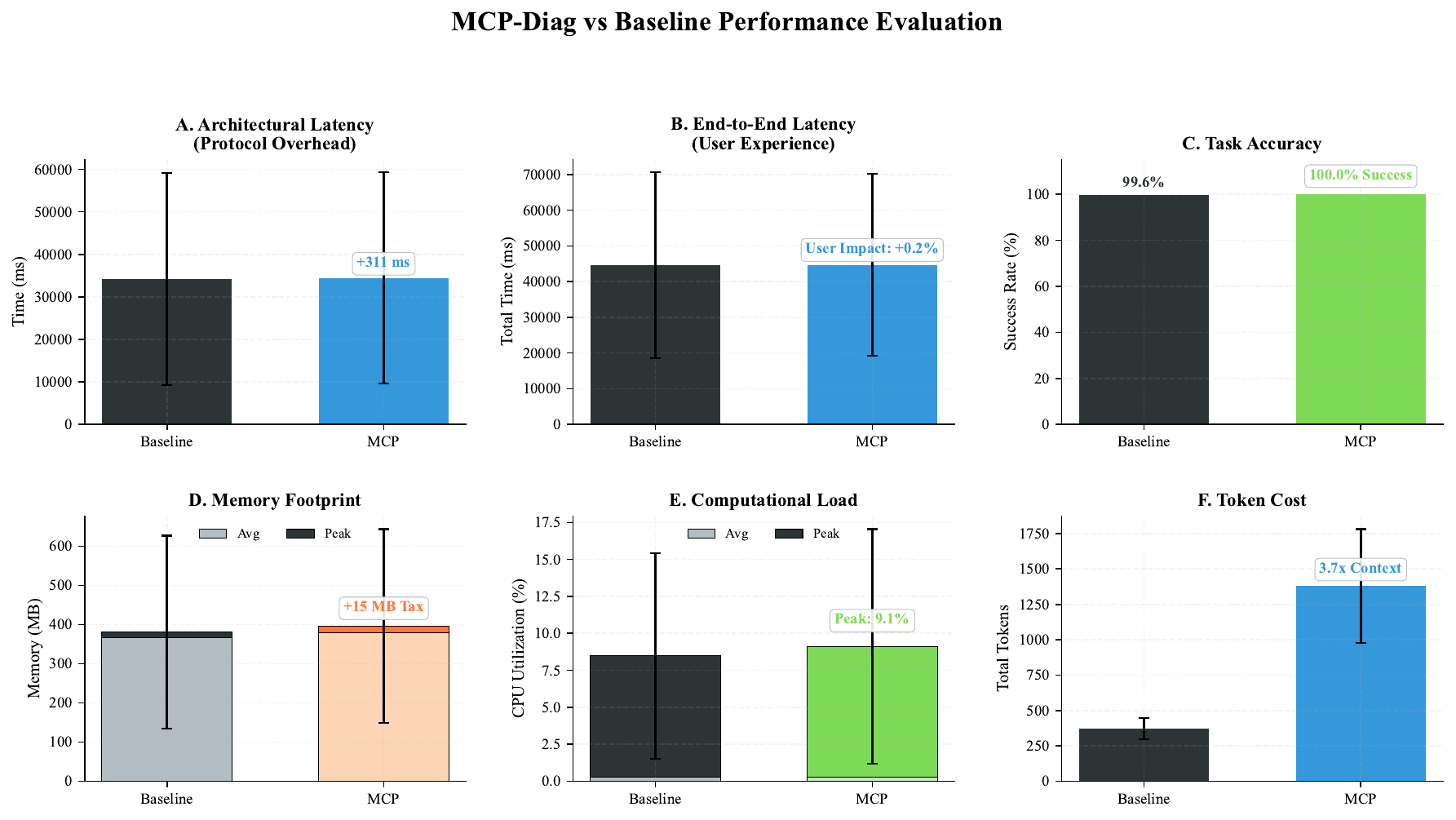}
    \caption{Comparison of MCP-Diag and baseline ShellTool for A) Protocol overhead B) Latency overhead C) Task Accuracy D) Memory overhead E) CPU overhead and F) Token Overhead over 500 trials. %($N=500$ Trials). (A) Memory Overhead (B) CPU Overhead (C) Token Overhead.
    %The protocol introduces a negligible $<0.9\%$ latency overhead. (C) MCP achieves perfect reliability vs. the baseline's tail-end failures. (D, E) Resource overhead is minimal (+15MB RAM, +1.1\% CPU). (F) Reliability comes at the cost of a $3.7\times$ increase in token consumption due to schema definitions.
    }
    \label{fig:benchmark}
\end{figure*}

\subsection{Grounding Fidelity and Reliability}
Experimental data (Fig. \ref{fig:benchmark}C) indicates that the \textit{ShellTool} baseline is inherently unstable, exhibiting stochastic failures in 0.4\% of trials. These errors occurred when non-standard vendor \texttt{stdout} formatting (e.g., unexpected ICMP error codes or null-hop asterisks) exceeded the regex-based parser's capabilities. By enforcing a rigid, type-safe JSON contract, MCP-Diag achieved 100\% extraction accuracy across the entire dataset.

\subsection{Architectural Tax: Latency and Token Cost}
We quantify the ``Architectural Tax'' as the overhead introduced by the protocol layer:
\begin{enumerate}
    \item \textbf{Latency Overhead:} As shown in Fig. \ref{fig:benchmark}A, the MCP layer introduces a mean overhead of 311ms. In the context of the traceroute task (avg. 34s), this tax is $<0.9\%$, resulting in a negligible user experience impact of $+0.2\%$ (Fig. \ref{fig:benchmark}B).
    \item \textbf{Token Consumption:} Enforcing JSON schemas results in a $3.7\times$ increase in context token usage per turn (mean 1100 vs. 300, Fig. \ref{fig:benchmark}F). Because the LLM processes input tokens in parallel, this increase does not significantly degrade time-to-first-token (TTFT) latency.
\end{enumerate}

\subsection{Resource Efficiency}
Fig. \ref{fig:benchmark}D, E show that the persistent MCP sidecar imposes a memory tax of only +15 MB and a peak CPU increase of +1.1\%. This demonstrates that the benefits of deterministic grounding can be achieved with a minimal footprint suitable for resource-constrained edge deployments.
\section{Conclusion and Future Work}
\label{sec:conclusion}
This paper introduced MCP-Diag, a novel architecture that provides a principled, security-first solution to the critical ``Security Gap" and ``Translation Gap" hindering the application of LLMs to network diagnostics. By synthesizing mandatory protocol-level elicitation, deterministic server-side data transformation, and a hybrid communication model, our work establishes a practical, production-ready blueprint for the future of AIOps.
The architectural foundation of MCP-Diag enables several ambitious avenues for future research:

\textbf{1. Complexity %Scaling % with Tool Complexity
:} As diagnostic tools evolve to capture deeper system states (e.g., full packet captures via \texttt{tshark}), the richness and depth of the structured JSON output will naturally increase. We anticipate that rapid advancements in LLM context windows and the development of more intelligent, adaptive parsers will seamlessly accommodate this growth, allowing for increasingly granular diagnostics without compromising the semantic rigor that our architecture guarantees.

\textbf{2. Autonomous Diagnostic Agents:} 
%We will leverage this expanded toolset to build a truly \textbf{autonomous diagnostic agent}. This involves 
We plan to utilize MCP's \texttt{Prompt} primitive to create complex, multi-step reasoning chains. With this approach, agent can autonomously diagnose a high-level symptom by sequencing \texttt{ping}, \texttt{traceroute}, and \texttt{dnsLookup} calls, while adhering to mandatory user approval at critical state transitions.

\textbf{3. Scalability and Concurrency:} Our current study focuses on the fundamental interaction strictly between a single MCP Host and Server. In future, we will explore the dimensions of concurrent execution and scalability. We will also investigate a distributed orchestration layer capable of managing thousands of concurrent MCP-Diag instances across a network fabric, transforming the architecture from a local diagnostic utility into a global sensor network.
%\vspace{-3mm}
\section*{Acknowledgment}
We thank the TPC Chairs and anonymous reviewers for their thoughtful feedback. 
This work is supported by grants from The Govt. of India, Department of science and technology, SERB Core Research Grant: CRG/2023/008021, MeitY Grant: 4(3)/2024-ITEA and Department of Telecommunications (DoT) 5G Use Case Lab, IIT Gandhinagar.

% --- REFERENCES ---
% pin references to the last page.
%\clearpage
%\raggedbottom
\balance
\bibliographystyle{IEEEtran}
\bibliography{references}

\begin{thebibliography}{10}
\providecommand{\url}[1]{#1}
\csname url@samestyle\endcsname
\providecommand{\newblock}{\relax}
\providecommand{\bibinfo}[2]{#2}
\providecommand{\BIBentrySTDinterwordspacing}{\spaceskip=0pt\relax}
\providecommand{\BIBentryALTinterwordstretchfactor}{4}
\providecommand{\BIBentryALTinterwordspacing}{\spaceskip=\fontdimen2\font plus
\BIBentryALTinterwordstretchfactor\fontdimen3\font minus \fontdimen4\font\relax}
\providecommand{\BIBforeignlanguage}[2]{{%
\expandafter\ifx\csname l@#1\endcsname\relax
\typeout{** WARNING: IEEEtran.bst: No hyphenation pattern has been}%
\typeout{** loaded for the language `#1'. Using the pattern for}%
\typeout{** the default language instead.}%
\else
\language=\csname l@#1\endcsname
\fi
#2}}
\providecommand{\BIBdecl}{\relax}
\BIBdecl

\bibitem{sdn_challenges_2024}
\BIBentryALTinterwordspacing
C.~Alice and H.~Bukola, ``Overcoming software-defined networking ({SDN}) challenges: Emerging solutions and future research directions,'' \emph{EasyChair Preprint}, no. 14687, 2024, easyChair Preprint no. 14687. [Online]. Available: \url{https://easychair.org/publications/preprint/pMvmj}
\BIBentrySTDinterwordspacing

\bibitem{edelman2018network}
J.~Edelman, S.~S. Lowe, and M.~Oswalt, \emph{Network Programmability and Automation: Skills for the Next-Generation Network Engineer}.\hskip 1em plus 0.5em minus 0.4em\relax Sebastopol, CA: O'Reilly Media, 2018.

\bibitem{rfc3535}
\BIBentryALTinterwordspacing
J.~Schoenwaelder, ``Overview of the 2002 {IAB} network management workshop,'' Internet Requests for Comments, RFC Editor, RFC 3535, May 2003. [Online]. Available: \url{https://www.rfc-editor.org/rfc/rfc3535.txt}
\BIBentrySTDinterwordspacing

\bibitem{zhang2025survey}
L.~Zhang, T.~Jia, Z.~Liu, Y.~Feng, W.~Zeng, and Y.~Yang, ``A survey of aiops in the era of large language models,'' \emph{ACM Computing Surveys}, 2025, just Accepted.

\bibitem{liu2024large}
\BIBentryALTinterwordspacing
C.~Liu, X.~Xie, X.~Zhang, and Y.~Cui, ``Large language models for networking: Workflow, advances and challenges,'' 2024. [Online]. Available: \url{https://arxiv.org/abs/2404.12901}
\BIBentrySTDinterwordspacing

\bibitem{jc_library}
\BIBentryALTinterwordspacing
K.~Brazil, ``{jc: CLI tool and python library that converts the output of popular command-line tools to JSON},'' 2024, gitHub Repository. Accessed: Dec. 20, 2025. [Online]. Available: \url{https://github.com/kellyjonbrazil/jc}
\BIBentrySTDinterwordspacing

\bibitem{hong2025comprehensive}
\BIBentryALTinterwordspacing
J.~Hong, N.~V. Tu, and J.~W.-K. Hong, ``A comprehensive survey on {LLM}-based network management and operations,'' \emph{International Journal of Network Management}, vol.~35, no.~6, 2025. [Online]. Available: \url{https://doi.org/10.1002/nem.70029}
\BIBentrySTDinterwordspacing

\bibitem{owasp_llm_top10_2025}
\BIBentryALTinterwordspacing
{OWASP Foundation}. (2025) {OWASP Top 10 for Large Language Model Applications (v2025)}. Specifies Vulnerability LLM06: Excessive Agency. Accessed: Dec. 20, 2025. [Online]. Available: \url{https://owasp.org/www-project-top-10-for-large-language-model-applications/}
\BIBentrySTDinterwordspacing

\bibitem{cve_2025_1753}
\BIBentryALTinterwordspacing
{National Vulnerability Database}. (2025, May) {CVE-2025-1753: Arbitrary Shell Execution via Argument Injection}. Accessed: Dec. 20, 2025. [Online]. Available: \url{https://nvd.nist.gov/vuln/detail/CVE-2025-1753}
\BIBentrySTDinterwordspacing

\bibitem{mcp_spec}
\BIBentryALTinterwordspacing
{Agentic AI Foundation}. (2025) {Model Context Protocol Specification}. V2025.11.25. Accessed: Dec. 20, 2025. [Online]. Available: \url{https://modelcontextprotocol.io/specification}
\BIBentrySTDinterwordspacing

\bibitem{langchain}
\BIBentryALTinterwordspacing
H.~Chase, ``{LangChain: Building applications with LLMs through composability},'' 2025, software Version 0.3. [Online]. Available: \url{https://github.com/langchain-ai/langchain}
\BIBentrySTDinterwordspacing

\bibitem{openinterpreter}
\BIBentryALTinterwordspacing
K.~Lucas, ``{Open Interpreter: A natural language interface for computers},'' 2023, gitHub Repository. [Online]. Available: \url{https://github.com/OpenInterpreter/open-interpreter}
\BIBentrySTDinterwordspacing

\bibitem{shellgpt}
\BIBentryALTinterwordspacing
E.~(TheR1D), ``{Shell-GPT: A command-line productivity tool powered by AI},'' 2023, gitHub Repository. [Online]. Available: \url{https://github.com/TheR1D/shell_gpt}
\BIBentrySTDinterwordspacing

\bibitem{gorilla_cli}
\BIBentryALTinterwordspacing
S.~G. Patil, T.~Zhang, X.~Wang, and J.~E. Gonzalez, ``{Gorilla: Large Language Model Connected with Massive APIs},'' 2023. [Online]. Available: \url{https://arxiv.org/abs/2305.15334}
\BIBentrySTDinterwordspacing

\bibitem{mcp_nettools_github}
\BIBentryALTinterwordspacing
D.~Kruyt, ``{mcp-nettools: A Model Context Protocol implementation providing network diagnostics},'' 2025, gitHub Repository. [Online]. Available: \url{https://github.com/dkruyt/mcp-nettools}
\BIBentrySTDinterwordspacing

\bibitem{warp_ai}
\BIBentryALTinterwordspacing
{Warp Technologies, Inc.} (2025) {Warp: The terminal for the 21st century}. [Online]. Available: \url{https://www.warp.dev}
\BIBentrySTDinterwordspacing

\bibitem{anthropic_computer_use}
\BIBentryALTinterwordspacing
{Anthropic}. (2024, Oct.) {Developing a computer use model}. [Online]. Available: \url{https://www.anthropic.com/news/developing-computer-use}
\BIBentrySTDinterwordspacing

\bibitem{mcp_diag_repo}
D.~Lodha, M.~Panchal, and S.~G. Kulkarni, ``{MCP-Diag: A Deterministic, Protocol-Driven Architecture for AI-Native Network Diagnostics},'' \url{https://github.com/devansh-lodha/mcp-diag}, 2025.

\bibitem{moz_top500}
\BIBentryALTinterwordspacing
{Moz}. (2025) {The Moz Top 500 Websites}. [Online]. Available: \url{https://moz.com/top500}
\BIBentrySTDinterwordspacing

\end{thebibliography}

\end{document}